# Full Analytical 3D Ion Transport Model for Large Deformable IPMC Soft Actuator

Mohsen Annabestani, Nadia Naghavi, and Mohammad Maymandi Nejad

*Abstract --* **Ionic Polymer Metal Composite is a well-known soft electroactive polymer composite that it's promising features tell us that it has adequate potential to be a utilizable and applicative soft actuator in the practical applications, especially in the small size applications. But this smart material is still immature, and one of the reasons that lead to its immaturity is lack of a valid and proper physics-based model for large deformation situations. In practical online and large deformation applications, the inverse non-autoregressive identification based models are the proper choices but if we want to know how IPMC works and what is the physics of its behavior in the large deformation situation the inverse identification based models are deeply blind, and we have to use physical and multi-physical approaches. It is our main aim in this paper, and for the first time, we want to present a fully analytical and physics-based ion transport 3D and non-Linear model for large deformable IPMC. In this direction, based on three dimensional Nernst-Plank PDE we will find a well-defined and valid relationship between input voltage and output tip displacement of IPMC for large deformation situation and with four provable pieces of evidence it will be proven that proposed model has chosen a proper way and it is more complete than previous benchmark and well-known physics-based models for IPMC, and also it is valid and accurate enough for large deformation modeling of IPMC.**

## NOMENCLATURES

| | |
|---|---|
| $h$ | Half the thickness of the Nafion membrane |
| $W$ | Width of the Nafion membrane |
| $L$ | Length of the Nafion membrane |
| $t_h$ | The thickness of the Pt electrode |
| $D(x,y,z,t)$ | Electrical displacement |
| $E(x,y,z,t)$ | Electric field |
| $\phi(x,y,z,t)$ | Electric potential |
| $\rho(x,y,z,t)$ | Electric charge density |
| $\kappa_e$ | The effective dielectric constant of Nafion |
| $F$ | Faraday constant |
| $C^+(x,y,z,t)$ | Concentration of cations |
| $C^-$ | Concentration of anions |
| $J(x,y,z,t)$ | Spatial vector of the cations |
| $d$ | Ionic diffusivity |
| $R$ | Gas constant |
| $T$ | Absolute temperature |
| $p$ | Fluid pressure |
| $\mathbf{v}$ | The free solvent velocity field |
| $\Delta V$ | Volumetric change of the Nafion |
| $k'$ | Hydraulic permeability coefficient |
| $V_{if}(t)$ | Input applied voltage to IPMC |
| $\sigma(x,y,z,t)$ | Mechanical stress |
| $\sigma_0$ | Constant coupling coefficient |
| $M(x,y,z,t)$ | Global bending moment |
| $\mathbf{r}$ | Position Vector |
| $M_{\Omega}(z,t)$ | Effective bending moment for IPMC |
| $Y$ | Young's modulus of Nafion |
| $I$ | Moment of inertia |
| $\omega(z,t)$ | Curvature shape function of IPMC |
| $\kappa_{Math}(z,t)$ | The curvature of $\omega(z,t)$ (Mathematical definition) |
| $\kappa_{Mech}(z,t)$ | The curvature of $\omega(z,t)$ (Mechanical definition) |
| $\delta_{Tip}(t)$ | Tip displacement function of IPMC |

## I. INTRODUCTION

Soft-robotics is a new promising subfield of robotics that its main target is constructing robots using soft materials in order to make more flexible, more bio-inspired, and safer robots. One of the serious choices as a soft actuator is Ionic polymer-metal composites (IPMCs). Due to particular properties of IPMC such as very large stimulus strain, low density, high toughness and lightness [1-7], this soft actuator is more suitable for micro and small size bioapplications. IPMC has a thin polymer membrane (usually Nafion) that is coated by two noble metal electrodes (usually Pt) on both sides (Anod a cathode). When we apply a low voltage (less than 5 V) to IPMC, it will be bent toward the anode side. Inversely, a low voltage will be generated between the two electrodes in response to bending of IPMC [5]. Hence we can say that IPMC is an actuator as well as sensors [1-5, 8-11]. We can find that in the vast variety of application we need largely bendable IPMCs but IPMS's behavior in large deformation situations is nonlinear [5, 12] and besides that as it has been mentioned in the [5, 12, 13], acquiring output feedback is not feasible in dominant practical applications of IPMCs. Hence it is a sensible expectation that we should choose a non-autoregressive (non-feedback) and large bendable approach in the modeling of IPMC, and we will choose and propose it in this paper.

Generally, we can divide IPMC's modeling approaches into two groups, the first group is analytical modeling, and the second one is predictive identification methods. Analytical models like using distributed Resistor-Capacitor (RC) equivalent circuits [13-19]. Or physical and multi-physical approaches using PDEs, Finite element methods, COMSOL modeling, etc [20-23]. And also there are some other miscellaneous uncategorized analytical models like [24-26]. In the predictive identification paradigms also we have two approaches, classical and intelligent methods. In classical methods, all papers except [5] have used autoregressive structures [27-30]. In intelligent methods also, all are autoregressive [2, 3, 31, 32] and except [2] and [3] are not valid for large deformation situation. Hence we can find that all mentioned methods except [5] in two above main categories are invalid for large deformation situations or they have used autoregressive or feed-backed structures. It means that paper [5] is the only paper that has been presented a model for identification of large deformation behavior of IPMC using a non-autoregressive method, but it is also completely black, and it cannot tells us about physics of IPMC. In any way if our main aim is merely using output of the model in practical applications this kind of identification methods like [5] are the proper choices but if we want to know how IPMC works and what is the physics of its behavior in the large deformation situation the models like [5] are deeply blind and we have to use physical and multi-physical approaches. It is our main aim in this paper, and we want to obtain a fully analytical and physics-based ion transport 3D non-Linear and feed-forward model for large deformable IPMC.

As another classification for physical and multi-physical ion and water transport models for IPMC, we can classify these models into three categories, thermodynamics of irreversible process (TIP) models, frictional models (FR) and Nernst-Planck PDE based (NP) models [33]. The key solution of TIP theory in the ion transport-based modeling of IPMC is that we can model the mass transport process based on the assumption of local equilibrium in this theory [33]. Based on TIP theory De Gennes et al. [34] presented a transport model for ion and water molecules. After Gennes some other similar approaches were also developed by Shahinpoor and Bar-Cohen [1], Paquette et al. [35], and Newbury and Leo [36]. FR model works differently, in this category it has assumed that at steady state the driving forces are balanced by frictional interactions among various components in IPMCs [33]. This concept is the governing rule in the transport process of ions and water molecules in the IPMC. Tadokoro et al. in 2000 proposed the first FR model [37], after that other researchers like Gong[26], Toi [38], Branco [39] and Yamaue [40] presented an improved version of Tadokoro model. In these three groups, NP models propose the most straightforward way to explain ion transport [33] that this approach is the most compatible approach with the nature of operation of ionic content into IPMC. The first time NP PDE was presented for charge distribution in an IPMC by Nemat-Nasser [9] and then improved his model in 2002[8], after that Farinholt [41, 42] used this method for representation of actuation and sensing response of IPMC and also for Modeling the electrical impedance response of IPMC [43]. This approach has been frequently used by other researchers, for example, Chen [44, 45] and Choonghee [46] have used NP PDEs for modeling of the electroactive deformation and sensing behavior of IPMC. NP PDEs have also been widely used by many others [24, 47-54] in it's modified, simplified, or incomplete version to show the process of cation migration in IPMC sensors and actuators. As mentioned before NP PDEs based models can propose the most straightforward solution to explain ion transport behavior of IPMC, we want to use and solve this PDE to model ion transport process of IPMC and then modeling of its curvature and mechanical deformation in response to the input voltage. We believe the ionic content of IPMC migrate through all directions and we need to solve this problem in the three dimensions. All above mentioned NP-based method except [38] are one dimensional, [38] also is a two dimensional, non-analytical and semi-numerical method. By the way we can say that our proposed method is the first dynamic full analytical 3D ion transport based model that can cover all linearity and non-linearity of IPMC's behavior even in large deformation situation and it is a promising result that for the first time we can find a mathematical well-defined relationship between input voltage and IPMC curvature in large deformation situation.

The rest of the paper consists of three parts, part II as the main part of this paper is about proposed 3D physics-based Ion transport model and its details. After that in part III, we will estimate the unknown parameters of our model and then assess its accuracy. Finally, in part IV, we will briefly talk about the achievements of our proposed model.

## II. 3D PHYSICS-BASED ION TRANSPORT MODEL FOR IPMC

### A. Preliminary equations and obtaining the governing PDE of electric potential

Let assume we have an IPMC like Fig.1:

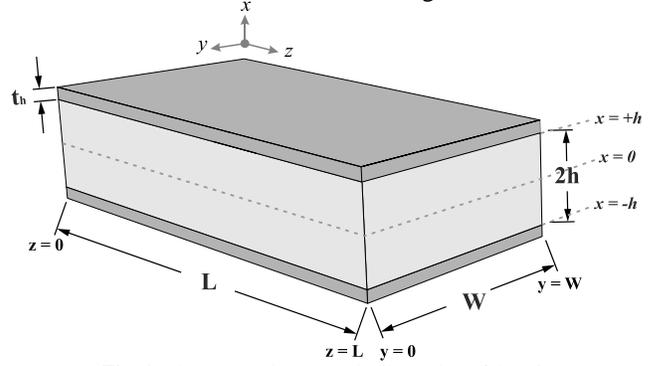

**Fig. 1.** The assumed parametric dimension of IPMC.

$\mathbf{D}(x,y,z,t)$ , $\mathbf{E}(x,y,z,t)$ , $\phi(x,y,z,t)$ and $\rho(x,y,z,t)$ represent electrical displacement, electric field, electric potential, and electric charge density, respectively. According to accepted principles in physics of electricity and magnetism, these components can be linked by the following relationships:

$$\mathbf{E}(x,y,z,t) = \frac{\mathbf{D}(x,y,z,t)}{\kappa_e} \tag{1}$$

$$\mathbf{E}(x,y,z,t) = -\nabla\phi(x,y,z,t) \tag{2}$$

$$\nabla.\mathbf{D}(x,y,z,t) = \rho(x,y,z,t) \tag{3}$$

Which $(x,y,z) \in [-h,+h] \times [0,W] \times [0,L]$ and $\kappa_e$ is dielectric constant and because we want to use these equations in describing the behavior of IPMC, this coefficient is the effective dielectric constant of Nafion membrane containing sodium ions, and $\rho(x,y,z,t)$ is also defined as follows[8, 9, 41, 42, 44, 45]:

$$\rho(x,y,z,t) = F\left(C^+(x,y,z,t) - C^-\right) \tag{4}$$

Where $F$ is the Faraday constant and $C^+(x,y,z,t)$ and $C^-$ are respectively the concentration of cations and anions in the membrane. Based on the continuity equation for cations, we also can find the following equation:

$$\nabla \cdot \mathbf{J}\left(x,y,z,t\right) = -\frac{\partial C^+\left(x,y,z,t\right)}{\partial t} \qquad (5)$$

$\mathbf{J}\left(x,y,z,t\right)$ is the spatial vector of the cations flux, and it is obtained from the solution of the following PDE, known as the Nernst-Planck PDE.

$$\mathbf{J} = -d\left(\nabla C^+ + \frac{C^+ F}{RT}\nabla\phi + \frac{C^+ \Delta V}{RT}\nabla p\right) + C^+ \mathbf{v} \qquad (6)$$

In the above equation, the phenomena of diffusion, ion migration, and convection have been considered that $d$ is the ionic diffusivity, $R$ is the gas constant, $T$ is the absolute temperature, $p$ is the fluid pressure, $\mathbf{v}$ is the free solvent velocity field and finally $\Delta V$ represents the volumetric change of the membrane[45].

***Note:*** *In this model, we solve the equations in a three dimensions time-variant space; that is, the equations governing the model are four variables, and they have $\left(x,y,z,t\right)$ mathematic argument. But from Eq (6), in order to avoid increasing the size of equations and save space, we avoid writing this argument. Unless in those equations that lack argument makes reader mislead.*

Also, with the help of the modified version of Darcy's law [55], we can link the free solvent velocity field with the gradient of fluid pressure and the electric field as follows[41, 42]:

$$\mathbf{v} = k'\left(C^- F\mathbf{E} - \nabla p\right) \qquad (7)$$

Where $k'$ is the hydraulic permeability coefficient. The second term of the right side of equation (6) is the term of convection, but since we know in the practical application of IPMC the gamut of temperature changes is not very varied and also we know the IPMC's performance is not very sensitive to temperature an especially in its optimum operation frequency (Less than 1 Hz) we can find that IPMC's performance is almost independent of temperature [56], hence convection term is ignorable, and we choose it equal to zero. To achieve zero convection term, since $c^+$ is the opposite of zero, we need to approach $\mathbf{v}$ to zero. If we approach $\mathbf{v}$ to zero, the fluid pressure gradient will be obtained according to the equation (8):

$$if \ \mathbf{v} \to 0 \Rightarrow \nabla p = C^- F\mathbf{E} \qquad (8)$$

Now by combination Eq (1) to (4), (6) and (8), we will have the following formula for $\boldsymbol{J}$ :

$$\mathbf{J} = -d\left(\frac{\kappa_e}{F}\nabla\left(\nabla\cdot\mathbf{E}\right) - \frac{\kappa_e}{RT}\left(1 - C^-\Delta V\right)\nabla\cdot\mathbf{E}\cdot\mathbf{E} - \frac{FC^-}{RT}\left(1 - C^-\Delta V\right)\mathbf{E}\right) \qquad (9)$$

In order to reach the $\nabla\cdot\mathbf{J}$, the first we apply the divergence operator ($\nabla\cdot$) in both sides of Eq (9), then based on Eqs (1) and (3) and fundamental principles of vector calculus we substitute $\kappa_e^{-1}\rho$ , $\nabla^2\rho$ and $\left(\nabla\rho\right)\cdot\mathbf{E} + \kappa_e^{-1}\rho^2$ instead of $\nabla\cdot\mathbf{E}$, $\nabla\cdot\left(\nabla\rho\right)$ and $\nabla\cdot\left(\rho\cdot\mathbf{E}\right)$ respectively. Hence we will have:

$$\nabla \cdot \mathbf{J} = -d\left(\frac{1}{F}\nabla\rho^2 - \underbrace{\tilde{K}\left(\left(\nabla\rho\right)\cdot\mathbf{E} + \frac{1}{\kappa_e}\rho^2\right)}_{term\ 1} - \underbrace{\frac{FC^-}{\kappa_e}\tilde{K}\rho}_{term\ 2}\right) \qquad (10)$$

Where $\tilde{K}$ is:

$$\tilde{K} = \frac{1}{RT}\left(1 - C^-\Delta V\right) \qquad (11)$$

In the equation (10) we can find that the ratio between the term 2 and term 1 is on the order of $5\times10^{-10}$, and it means that term 2 is too smaller than term 1 and so we can ignore term 2 [41], and as a result, Eq (10) becomes the following:

$$\nabla \cdot \mathbf{J} = -d\left(\frac{1}{F}\nabla\rho^2 - \frac{FC^-}{\kappa_e}\tilde{K}\rho\right) \qquad (12)$$

Then using a combination of equations (12), (5) and (4) we can find a PDE for calculation of $\rho$ as follows:

$$\frac{\partial\rho}{\partial t} = d\nabla^2\rho - K\rho \qquad (13)$$

Where $K$ is $F^2C^-d\kappa_e^{-1}\tilde{K}$ and defined as (14):

$$K = \frac{F^2C^-d\left(1 - C^-\Delta V\right)}{RT\kappa_e} \qquad (14)$$

Now if we combine Eqs (1) to (3) the following relations will be obtained between electric potential and electric charge density:

$$\rho = -\kappa_e\nabla^2\phi \qquad (15)$$

And finally, if we substitute (15) in (13) the following time variant 3D PDE will be resulted for $\phi\left(x,y,z,t\right)$ . In the next part, we want to solve this PDE.

$$\frac{\partial}{\partial t}\left(\nabla^2\phi\right) = d\nabla^2\left(\nabla^2\phi\right) - K\left(\nabla^2\phi\right) \qquad (16)$$

***Evidence 1:*** The above equation is the more complete version of PDE that Nemat-Nasser first time found for $\phi\left(x,t\right)$ in 2000 [9]. If we extend Eq (16) just for $x$ variable, easily it will be changed to (17) that this equation is exactly the same PDE that Nemat-Nasser found for $\phi\left(x,t\right)$ (Please see Eq 37 of [9]).

$$\frac{\partial^3\phi}{\partial x^2\partial t} = d\left(\frac{\partial^4\phi}{\partial x^4} - a^2\frac{\partial^2\phi}{\partial x^2}\right) \qquad (17)$$

Where $a$ is:

$$a = \sqrt{\frac{F^2C^-\left(1 - C^-\Delta V\right)}{RT\kappa_e}} \qquad (18)$$

As the first evidence, it can assure us that we have chosen a proper and more common way for our model.

### B. Calculation of $\phi\left(x,y,z,t\right)$

The answer of Eq (16) is the four-variable scalar function $\phi\left(x,y,z,t\right)$ .To solve this PDE, we need proper boundary conditions. The $\phi\left(x,y,z,t\right)$ represents the electric potential of the IPMC, and it is obvious that the electric potential of the IPMC in the clamping area on the surface of the electrodes is the same IPMC applied input voltage, which we call it $V_I(t)$. Thus, the governing boundary conditions of this boundary problem will be expressed as (19).

$$\begin{cases} \phi(+h,0,0,t)=\dfrac{V_I(t)}{2} \\ \phi(-h,0,0,t)=-\dfrac{V_I(t)}{2} \\ \phi(+h,0,L,t)=a_L(V_I(t))\dfrac{V_I(t)}{2} \\ \phi(-h,0,L,t)=-a_L(V_I(t))\dfrac{V_I(t)}{2} \end{cases} \begin{cases} \phi(+h,W,0,t)=\dfrac{V_I(t)}{2} \\ \phi(-h,W,0,t)=-\dfrac{V_I(t)}{2} \\ \phi(+h,W,L,t)=a_L(V_I(t))\dfrac{V_I(t)}{2} \\ \phi(-h,W,L,t)=-a_L(V_I(t))\dfrac{V_I(t)}{2} \end{cases} \quad (19)$$

In these conditions, the $a_L(V_I(t))$ function is called longitudinal surface voltage attenuation function. Ideally, if IPMC electrodes are considered as perfect conductors and the fractal penetration of the electrodes into the membrane is ignored, the voltage of the clamp region is constant across the whole surface of electrodes, and there will be no any attenuations in its longitudinal direction [26]. However, since that the electrodes do not have the conductivity of a perfect conductor and also due to electrode deposition process and fractal penetration of electrodes inside the membrane, the applied voltage will be attenuated along the $z$-axis from clamping area of IPMC proportional to amplitude, frequency and other components of the input voltage. We describe this attenuation with a coefficient derived from $V_I(t)$ as $a_L(V_I(t))$ function. This function is a continuous function that produces an integer in the interval $(0,1]$, which 1 means that the electrodes have perfect conductivity and means that we haven't considered the surface resistance. As a result, the measured voltage at the IPMC's tip has not been attenuated and is equal to the voltage of clamp area. Naturally, if this number is less than 1, means that we have seen the longitudinal attenuation of voltage in the problem and consider the surface resistance, which this state is more like reality.

The method that we have chosen to solve Eq (16) is the separation of variables method that the calculated $\phi(x,y,z,t)$ using this method is:

$$\phi(x,y,z,t)=k_1 e^{-\lambda t}\left[U_1(x,y,z,t)+U_2(x,y,z)\right] \quad (20)$$

Where $U_1(x,y,z,t)$ and $U_2(x,y,z)$ are defined as (21) and (22):

$$U_1(x,y,z,t)=\left\{\dfrac{V_I(t)}{2}\left(\dfrac{Sinh(\hat{\alpha}_1(t)x)}{Sinh(\hat{\alpha}_1(t)h)}\right)\times\right.$$
$$\times\left(\left(\dfrac{1-\cosh(\hat{\alpha}_2 W)}{Sinh(\hat{\alpha}_2 W)}\right)Sinh(\hat{\alpha}_2 y)+\cosh(\hat{\alpha}_2 y)\right)\times \quad (21)$$
$$\times\left(\dfrac{a_L(V_I(t))-\cosh(\hat{\alpha}_3(t)L)}{sinh(\hat{\alpha}_3(t)L)}\right)\times$$
$$\left.\times\left(sinh(\hat{\alpha}_3(t)z)+cosh(\hat{\alpha}_3(t)z)\right)\right\}$$

$$U_2(x,y,z)=\sum_{m=0}^{\infty}\sum_{n=0}^{\infty}\sum_{p=0}^{\infty}a_{mnp}\ \Xi(x,y,z) \quad (22)$$

$\ni\ \Xi(x,y,z)=\sin\dfrac{(m+1)\pi}{2h}x\ \sin\dfrac{(n+1)\pi}{W}y\ \sin\dfrac{(p+1)\pi}{L}z$

$k_1$ and $\lambda$ also are constant coefficients and $\hat{\alpha}_1(t),\hat{\alpha}_2$ and $\hat{\alpha}_3(t)$ define as follows:

$$\hat{\alpha}_1(t)=\sqrt{\hat{K}-\left\{\dfrac{\mu^2}{W^2}\left(Ln\left(1-\dfrac{1}{\mu^2}\right)\right)^2+\dfrac{2\theta}{L}Ln\left(\dfrac{4}{a_L(V_I(t))L\theta}\right)\right\}} \quad (23)$$

$$\hat{\alpha}_2=\dfrac{\mu}{W}Ln\left(1-\dfrac{1}{\mu^2}\right) \quad (24)$$

$$\hat{\alpha}_3(t)=\sqrt{\dfrac{2\theta}{L}Ln\left(\dfrac{4}{a_L(V_I(t))L\theta}\right)} \quad (25)$$

Where $\theta$ is $\dfrac{1}{W}$ and $\mu$ is a constant coefficient that it should be bigger than 1. And also $\hat{\alpha}_1(t),\hat{\alpha}_2,\hat{\alpha}_3(t)$ have to always satisfy the following condition.

$$\hat{\alpha}_1(t)^2+\hat{\alpha}_2^2+\hat{\alpha}_3(t)^2=\hat{K} \quad (26)$$

$a_{mnp}$ also, define by (27):

$$a_{mnp}=\dfrac{\dfrac{8\hat{K}}{2hWL}\int_{-h}^{+h}\int_0^W\int_0^L\tilde{\Psi}(x,y,z)\,dx\,dy\,dz}{\left(\left(\dfrac{(m+1)\pi}{2h}\right)^2+\left(\dfrac{(n+1)\pi}{W}\right)^2+\left(\dfrac{(p+1)\pi}{L}\right)^2+\hat{K}\right)} \quad (27)$$

Where

$$\tilde{\Psi}(x,y,z)=\Psi(x,y,z)\times$$
$$\times\left(\sin\dfrac{(m+1)\pi}{2h}x\ \sin\dfrac{(n+1)\pi}{W}y\ \sin\dfrac{(p+1)\pi}{L}z\right) \quad (28)$$

And

$$\Psi(x,y,z)=\Upsilon_1 x+\Upsilon_2 y+\Upsilon_3 z+\Upsilon_4 xy+\Upsilon_5 xz+$$
$$+\Upsilon_6 yz+\Upsilon_7 xyz+\Upsilon_8+\dfrac{\gamma_1}{2}x^2+\dfrac{\gamma_2}{2}y^2+\dfrac{\gamma_3}{2}xz^2 \quad (29)$$

Where $\Upsilon_1$ to $\Upsilon_8$ and $\gamma_1$ to $\gamma_3$ are constant coefficients.

**Evidence 2:** If we extend Eq 20 just for small amount of $x$ variable (Means small deformation situation) and ignore y and z variables, it will be changed to the following equation that it is the exact equation that Farinholt [41, 42] found for electric potential (Please see Eq 2.42 of [41]).

$$\phi(x,t) = \frac{\rho_0}{\kappa_e \hat{\alpha}_1^{\ 2} \sinh(\hat{\alpha}_1 h)} \left( \sinh(\hat{\alpha}_1 x) - \frac{x}{h} \sinh(\hat{\alpha}_1 h) \right) e^{-\lambda t} \quad (30)$$

Where $\rho_0$ is the boundary value of $\rho(x,t)$ in $x=h$. As the second evidence, it can assure us again that we have chosen a proper way for our model and proves that the Farinholt approach is just a special state of our model that only valid for small deformation situations.

### C. Vector proper physics matched coupling relation

When the electrical voltage is applied to IPMC, the hydrated sodium cations, ions that a couple of water molecules bind to them, move toward the IPMC cathode side, which causes the concentration of water molecules to increase gradually there then leads to Nafion swelling at the cathode side. The result of this transport process is induced stress that causes mechanical bending of IPMC [26, 37]. Thus, in fact, migration and concentration of cations cause mechanical stress and bending of the IPMC. Hence we must first calculate the concentration of cations (Electrochemical component), and then, by defining a coupling relation corresponding to the physics of the problem, we obtain the stress (Mechanical component). The process that has been used so far in many articles [14, 41-45, 57, 58] is using the idea that has been proposed by Nemat Nasser [9]. The Nemat Naser's idea process is that they first obtain the electric charge density ($\rho$), then calculate the mechanical stress ($\sigma$) using $\sigma = \alpha \rho$ linear relationship. Which $\alpha$ is a constant coefficient and plays the role of coupling between charge density and mechanical stress. According to the Eq (4), the charge density and the concentration of cations have linear relationships with each other. Therefore, it can be said that according to the coupling relationship of Nemat Naser, the relationship between the cationic concentration (The electrochemical factor that makes swelling in IPMC's membrane) and mechanical stress (The mechanical factor that makes swelling in IPMC's membrane) is followed by the following relation, which it is a linear relationship.

$$\rho(x,t) = FC^+(x,t) - FC^- \xrightarrow{\sigma(x,t) = \alpha\rho(x,t)}$$
$$\rightarrow \sigma(x,t) = (\alpha F)C^+(x,t) - (\alpha FC^-) \quad (31)$$

Mathematically, this relation in the Nemat Nasser approach means that the behavior of concentration of cations ($C^+$) is similar to the behavior of mechanical stress ($\sigma$), just with a difference in their magnitudes, while the physics of the problem tells us the opposite of this similarity. Because that base on governing mechanics in a cantilever beam (IPMC here) for making a bending in the direction of the $x$-axis (Main direction of IPMC's bending) we need to have one component of induced stress in the direction of $z$-axis (Fig.2). On the other hand, we know the concentration of cations in the Nemat Nasser model is a function of $x$ and $t$ variables, and it makes an ambiguity because the stress function should have a $z$ variable and it is in the opposite of Eq (31). Hence we can say that this relation is not a proper coupling relation and is not match with the physics of the problem at least for large deformation situations of IPMC, and we should find a new proper physics-based coupling relation.

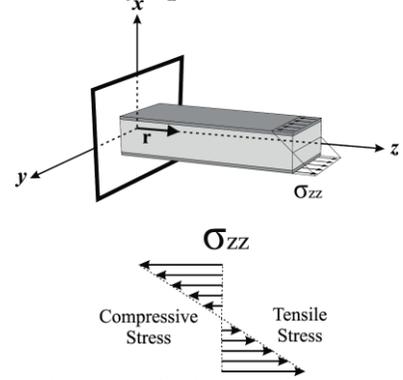

**Fig. 2.** For making a bending in the direction of $x$-axis (Main direction of IPMC's bending) we need to have induced stress in direction of $z$-axis

As we mentioned before the most of papers that have presented a physical model for IPMC have assumed that cations move in the direction of $x$-axis between anode and cathode and it means that they have assumed the concentration of cations are one dimensional and defined as a function of $x$ and $t$ variables [8, 9, 14, 17, 20, 21, 37, 41-45, 57-64]. But nature is different, and ions move to all directions (Of course the dominant of them move in the direction of $x$-axis). This ionic movement is depended on ionic conductivity, and ionic conductivity indirectly is depended on electrical conductivity, and electrical conductivity is inversely depended on electrical resistance. Hence instead of the assumption that says ions move just in the direction of $x$-axis, it would be better that we assume that ions move toward a spatial vector like $\mathbf{A}$ that the components of this vector are depended on electric resistance of IPMC in all axis that defined as follows :

$$\mathbf{A} = \Omega_x(r_M,t)\mathbf{i} + \Omega_y(r_{ew},t)\mathbf{j} + \Omega_z(r_{el},t)\mathbf{k} \quad (32)$$

Where $\Omega_x(r_M,t)$, $\Omega_y(r_{ew},t)$ and $\Omega_z(r_{el},t)$ are the functions of $r_M$, $r_{ew}$ and $r_{el}$, and $r_M$, $r_{ew}$ and $r_{el}$ are membrane resistance and transverse and longitudinal resistances of electrodes respectively. As we were saying, migration and changing the concentration indirectly make mechanical stress and the bending of IPMC, now we know that this changing of the concentration is done in direction of vector $\mathbf{A}$ and naturally $\Omega_x(r_M,t)$, $\Omega_y(r_{ew},t)$ and $\Omega_z(r_{el},t)$ determine the optimum direction of $\mathbf{A}$. Mathematically when we want to calculate the changes of multivariable function in direction of a vector, we should calculate the directional derivative of the function in direction of the desired vector. For example here since we want to find the changes of ion's concentration in direction of optimum movement's vector, we should calculate the directional derivative of $C^+(x,y,z,t)$ in direction of vector $\mathbf{A}$ that it is shown by $\nabla_{\mathbf{A}}C^+(x,y,z,t)$ and defines as follows:

$$\nabla_{\mathbf{A}}C^+ = \frac{\partial C^+}{\partial \mathbf{A}} = \mathbf{a}.\nabla C^+ \quad (33)$$

Where **a** is the unit vector of **A** and defined by $\mathbf{a} = \dfrac{\mathbf{A}}{|\mathbf{A}|}$. And $\nabla C^{+}$ is also the gradient of $C^{+}(x, y, z, t)$. We know that in the pure and basic IPMC (An IPMC without using any controller and any modification on it), the bending displacement is attenuated over time, and it is related to the effects of resistance. It means that the effects of resistance will be increased over time and make obstacles for current flow on electrodes and ion movement on the membrane. Hence it is sensible that we choose time-ascending functions for modeling the effects of resistance, the functions like the ramp, sigmoid, tangent hyperbolic, etc, because we want to increase the effects of resistance over time. Here we have chosen the ramp function, now if we use these ramp functions for $\Omega_x(r_M, t)$, $\Omega_y(r_{ew}, t)$ and $\Omega_z(r_{el}, t)$ like (34) to (36) and define mentioned resistances as the functions of dimensions (h, W, and L). We can calculate $\nabla_{\mathbf{A}} C^{+}$ as (37):

$$\Omega_x\left(r_M(h), t\right) = r_M(h) t\, u(\mathrm{t}) \tag{34}$$

$$\Omega_y\left(r_{ew}(W), t\right) = r_{ew}(W) t\, u(\mathrm{t}) \tag{35}$$

$$\Omega_z\left(r_{el}(L), t\right) = r_{el}(L) t\, u(\mathrm{t}) \tag{36}$$

$$\nabla_{\mathbf{A}} C^{+} = \frac{1}{\hat{r}(h, W, L)}\left(r_M(h)\frac{\partial C^{+}}{\partial x} + r_{ew}(W)\frac{\partial C^{+}}{\partial y} + r_{el}(L)\frac{\partial C^{+}}{\partial z}\right) \tag{37}$$

Where $u(t)$ is the Heaviside step function and $\hat{r}(h, W, L)$ is:

$$\hat{r}(h, W, L) = \sqrt{r_M(h)^2 + r_{ew}(W)^2 + r_{el}(L)^2} \tag{38}$$

By having $\nabla_{\mathbf{A}} C^{+}$ we can find the changes of ion's concentration in the direction of optimum movement's vector that these changes induce mechanical stress, semi-directly and proper physics matched. Hence we can say that there is a direct linear relationship like (39) between $\nabla_{\mathbf{A}} C^{+}$ and mechanical stress:

$$\sigma(x, y, z, t) = \sigma_0 \nabla_{\mathbf{A}} C^{+}(x, y, z, t) \tag{39}$$

Where $\sigma_0$ is a constant coupling coefficient that we should estimate it in the model estimation stage.

### D. Mechanical Stress

To find induced stress using Eq (39). The first we should calculate $\nabla_{\mathbf{A}} C^{+}$ and for calculation of $\nabla_{\mathbf{A}} C^{+}$ we need to know $C^{+}$. It is so simple to find $C^{+}$ by the combination of (1) to (4) as (40):

$$C^{+} = C^{-} - \frac{\kappa_e}{F}\nabla^2\phi \tag{40}$$

Now we can calculate $\nabla_{\mathbf{A}} C^{+}$ using (37) and finally $\sigma$ will be obtained by (41):

$$\sigma(x, y, z, t) = -\frac{\kappa_e \sigma_0}{F\,\hat{r}(h, W, L)}\left(A_1(x, y, z, t) + \sum_{i=2}^{4} A_i(x, y, z)\right) \tag{41}$$

Where

$$A_1 = \hat{K}\left(r_M(h)\frac{\partial U_1}{\partial x} + r_{ew}(W)\frac{\partial U_1}{\partial y} + r_{el}(L)\frac{\partial U_1}{\partial z}\right) \tag{42}$$

$$A_2 = r_M(h)\left(\frac{\partial^3 U_2}{\partial x^3} + \frac{\partial^3 U_2}{\partial x \partial y^2} + \frac{\partial^3 U_2}{\partial x \partial z^2}\right) \tag{43}$$

$$A_3 = r_{ew}(W)\left(\frac{\partial^3 U_2}{\partial y \partial x^2} + \frac{\partial^3 U_2}{\partial y^3} + \frac{\partial^3 U_2}{\partial y \partial z^2}\right) \tag{44}$$

$$A_4 = r_{el}(L)\left(\frac{\partial^3 U_2}{\partial z \partial x^2} + \frac{\partial^3 U_2}{\partial z \partial y^2} + \frac{\partial^3 U_2}{\partial z^3}\right) \tag{45}$$

Where $U_1$ and $U_2$ already were obtained by (21) and (22), and we can calculate $A_1$ to $A_4$ easily using (42) to (45).

### E. Bending moment

Based on beam theory, IPMC strip is a cantilever beam, and we can consider it as a cantilever beam to calculate it's bending moment. Basically, in a cantilever beam, the global bending moment is described by a vector like **M** that defined by (46):

$$\mathbf{M} = \iint_{c_y\ c_x} \mathbf{r} \times \left(\sigma_{zx}\,\mathbf{i} + \sigma_{zy}\,\mathbf{j} + \sigma_{zz}\,\mathbf{k}\right) dx\, dy \tag{46}$$

Where **r** is the position vector and is defined by $\mathbf{r} = x\mathbf{i} + y\mathbf{j}$. If we expand (46) we will have (47):

$$\mathbf{M} = M_{zx}\mathbf{i} + M_{zy}\mathbf{j} + M_{zz}\mathbf{k} \tag{47}$$

Where

$$M_{zx} = \iint_{c_y\ c_x} y\,\sigma_{zz}\ dx\, dy \tag{48}$$

$$M_{zy} = \iint_{c_y\ c_x} x\,\sigma_{zz}\ dx\, dy \tag{49}$$

$$M_{zz} = \iint_{c_y\ c_x}\left(x\,\sigma_{zy} - y\,\sigma_{zx}\right) dx\, dy \tag{50}$$

As has been depicted in Fig 3, $M_{zx}$ is absolute zero, and $M_{zz}$ is almost zero for IPMC, and in fact, only the $M_{zy}$ induces to the IPMC and makes IPMC bend in the direction of the x-axis. As described above we can say that the only effective bending moment for IPMC is $M_{zy}$ that if we extend it for the dimension of IPMC it could be defined as (51):

$$M_{zy}(z, t) = \int_0^W \int_{-h}^{+h} x\,\sigma_{zz}(x, y, z, t)\, dx\, dy \tag{51}$$

Where $\sigma_{zz}(x, y, z, t)$ is the renamed version of the same $\sigma(x, y, z, t)$ that we defined it in (41). Now by combination of (51) and (41) to (45) and solving integration terms and some mathematical efforts we can find $M_{zy}$ as (52):

$$M_{zy}(z, t) = -\frac{\kappa_e \sigma_0}{F\,\hat{r}(h, W, L)}\left(M_{1zy}(z, t) + M_{2zy}(z)\right) \tag{52}$$

Where

$$M_{1zy}(z, t) = -\left(\frac{2\,\hat{K}\,\hat{\alpha}_3(t)\, r_{el}(L)}{\hat{\alpha}_1(t)^2\,\hat{\alpha}_2}\right) \times$$
$$\times\left(1 - \hat{\alpha}_1(t)\coth\left(\hat{\alpha}_1(t)h\right)\right)\left(\tanh\left(\frac{\hat{\alpha}_2 W}{2}\right)\right) \times \tag{53}$$
$$\times\left(a_L\left(V_I(t)\right)\operatorname{csch}\left(\hat{\alpha}_3(t)L\right) - \coth\left(\hat{\alpha}_3(t)L\right)\right) \times$$
$$\times \cosh\left(\hat{\alpha}_3(t)z\right) V_I(t)$$

$$M_{2zy}(z) = -\left(\frac{2h^3 W\, b_{mnp}}{3}\right)\left(\hat{\Upsilon}_2 + \hat{\Upsilon}_5\frac{W}{2} + \hat{\Upsilon}_6 W z\right) \quad (54)$$

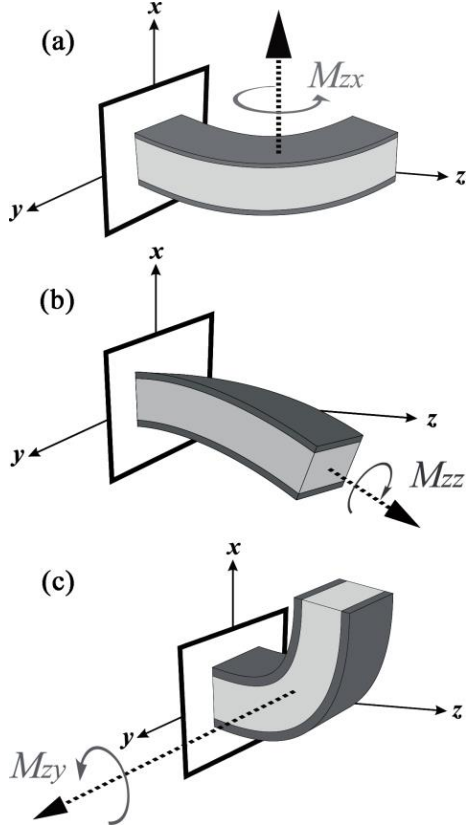

**(a)**

$M_{zx}$

**(b)**

$M_{zz}$

**(c)**

$M_{zy}$

**Fig. 3.** Three possible bending moments for the IPMC strip. $M_{zx}$ is absolute zero (a), and $M_{zz}$ is almost zero for IPMC (b). But $M_{zy}$ bending moment that makes IPMC bend in $x$-direction (c).

That $b_{mnp}$ is a constant coefficient that related to numbers of harmonics that we use in calculation of $U_2$ and $\hat{\Upsilon}_2$, $\hat{\Upsilon}_5$ and $\hat{\Upsilon}_6$ are constant coefficients that they have created by the combination of $\Upsilon_1$ to $\Upsilon_8$ and $\gamma_1$ to $\gamma_3$ and we should predict them in the model prediction stage. By the way, if we want to show $M_{zy}$ by a unit equation we can use the following:

$$M_{zy}(z,t) = \tilde{\Upsilon}_1(t) + \tilde{\Upsilon}_2(t)z + \tilde{\Upsilon}_3(t)\cosh\left(\hat{\alpha}_3(t)z\right) \quad (55)$$

Where $\tilde{\Upsilon}_1(t)$ to $\tilde{\Upsilon}_3(t)$ have been defined by (56) to (58) respectively:

$$\tilde{\Upsilon}_1(t) = \left(\frac{\kappa_e \sigma_0}{F\,\hat{r}\left(h, \mathrm{W}, \mathrm{L}\right)}t\right)\left(\frac{2h^3 W}{3}\right)\left(\hat{\Upsilon}_2 + \hat{\Upsilon}_5\frac{W}{2}\right)b_{mnp} \quad (56)$$

$$\tilde{\Upsilon}_2(t) = \left(\frac{\kappa_e \sigma_0}{F\,\hat{r}\left(h, \mathrm{W}, \mathrm{L}\right)}t\right)\left(\frac{2h^3 W}{3}\right)\left(\hat{\Upsilon}_6 W\right)b_{mnp} \quad (57)$$

$$\begin{aligned}
\tilde{\Upsilon}_3(t) = {} & \left(\frac{\kappa_e \sigma_0}{F\,\hat{r}\left(h, \mathrm{W}, \mathrm{L}\right)}t\right)\left(\frac{2\hat{K}\,\hat{\alpha}_3(t)\,r_{el}\left(L\right)}{\hat{\alpha}_1(t)^2\,\hat{\alpha}_2}\right)\times \\
& \times\left(1 - \hat{\alpha}_1(t)\coth\left(\hat{\alpha}_1(t)h\right)\right)\left(\tanh(\frac{\hat{\alpha}_2 W}{2})\right)\times \\
& \times\left(a_L\left(V_I(t)\right)\operatorname{csch}\left(\hat{\alpha}_3(t)L\right) - \coth\left(\hat{\alpha}_3(t)L\right)\right)V_I(t)
\end{aligned} \quad (58)$$

### F. Calculation of IPMC tip displacement in large deformation situation

To calculate the IPMC bending, the first we need to have curvature shape function of IPMC. Then using this function we can find its tip displacement. But in the most of paper that model IPMC behavior using physics-based approaches, there is a common mistake in this direction and it is applying linear Euler-Bernoulli (LEB) beam theory to calculate tip displacement of IPMC (For example [14, 41, 42, 44, 45, 57, 58, 65]). LEB beam theory only can consider the infinitesimal strains and it means that by this theory we can only model small deformation of a beam (Here IPMC) [66]. Based on LEB beam theory the relation of curvature shape function of IPMC, $\omega(z,t)$, and bending moment, $M_{zy}(z,t)$, is defined as follows:

$$\frac{\partial^2 \omega(z,t)}{\partial z^2} = \frac{M_{zy}(z,t)}{Y\,I} \quad (59)$$

Where $Y$ is Young's modulus of Nafion membrane and $I$ is moment of inertia and calculated by the following relation:

$$I = \int_0^L \int_0^W \int_{-h}^{+h}\left(x^2 + z^2\right)dx\,dy\,dz \Rightarrow I = \frac{2}{3}h^3 WL \quad (60)$$

As mentioned before, Eq (59) is valid just for infinitesimal strains, and we cannot use it for large deformation situations of IPMC. Hence we should find a valid relation for large deformation that this relation is main relation of curvature. Mathematically the curvature of a function like $\omega(z,t)$ will be calculated by $\kappa_{Math}(z,t)$ where defined as following:

$$\kappa_{Math}(z,t) = \frac{\dfrac{\partial^2 \omega(z,t)}{\partial z^2}}{\left(1 + \left(\dfrac{\partial \omega(z,t)}{\partial z}\right)^2\right)^{\frac{3}{2}}} \quad (61)$$

Mechanically also the curvature of a beam that $M_{zy}(z,t)$ bending moment has been induced on it is defined by (62):

$$\kappa_{Mech}(z,t) = \frac{M_{zy}(z,t)}{Y\,I} \quad (62)$$

Now our aim is the finding a dynamic function like $\omega(z,t)$ that its shape is match with the shape of a cantilever beam (Here IPMC) that a bending moment like $M_{zy}(z,t)$ has been induced on it. To find this function it is only enough to set $\kappa_{Math}(z,t)$ and $\kappa_{Mech}(z,t)$ equal and solve resulted PDE. If we do this, we will obtain following nonlinear PDE:

$$if \quad \kappa_{Math}(z,t) = \kappa_{Mech}(z,t) \Rightarrow$$

$$\Rightarrow \frac{\partial^2 \omega(z,t)}{\partial z^2} - \left( \frac{M_{zy}(z,t)}{Y I} \right) \left( 1 + \left( \frac{\partial \omega(z,t)}{\partial z} \right)^2 \right)^{\frac{3}{2}} = 0 \qquad (63)$$

***Evidence 3:*** In the above PDE, if we ignore term $\left( \frac{\partial \omega(z,t)}{\partial z} \right)^2$, it will be changed to governing PDE of LEB beam theory (Eq 59). Hence it will be the third evidence and assure us again that our method has chosen a proper way and has presented a complete approach to physics-based modeling of IPMC.

Eq (63) is a nonlinear PDE, but it is solvable analytically. If we solve this PDE, $\omega(z,t)$ will be obtained as follows:

$$\omega(z,t) = \frac{f(t) Ln\big(P(z,t)\big) + S(t) q(z,t)}{\left( \frac{f(t)}{g(t)} \right) Ln\left( \frac{1}{P(0,t)} \right) - S(t) q(0,t)} + g(t) \qquad (64)$$

Where:

$$P(z,t) = \left| \frac{f(t) \tan\left( \frac{q(z,t)}{2} \right) + S(t) - Y I}{f(t) \tan\left( \frac{q(z,t)}{2} \right) - S(t) - Y I} \right| \qquad (65)$$

$$S(t) = \sqrt{\big(YI - f(t)\big)\big(YI + f(t)\big)} \qquad (66)$$

$$q(z,t) = Arc \sin\left( \frac{m(z,t) + f(t)}{Y I} \right) \qquad (67)$$

$$f(t) = k_f \, e^{-\frac{t}{\tau_f}} \qquad (68)$$

$$g(t) = k_g \, e^{-\frac{t}{\tau_g}} \qquad (69)$$

That $k_f$, $k_g$, $\tau_f$ and $\tau_g$ are constant coefficients that they will be estimated in the model estimation stage.

$\omega(z,t)$ gives us the shape of IPMC, but we want to find the position of its tip overtime or the same dynamic tip displacement. If we name the function of IPMC tip displacement $\delta_{Tip}(t)$, it is obvious that the following equation, that has been used in the several papers like[45], is absolutely wrong, and it is just an approximation of IPMC tip displacement in the very small deformation situation.

$$\delta_{Tip}(t) = \omega(L,t) \qquad (70)$$

In the Eq (70), $L$ is the length of IPMC, and this relation is valid only when IPMC is open circuit and is completely off and don't have any movement. To solve this big mistake we should use the following formula:

$$\delta_{Tip}(t) = \omega\big(z_L(t),t\big) \qquad (71)$$

Where $z_L(t)$ makes the dynamic coordination of the $z$ component of IPMC's tip using Eq (72):

$$\int_0^{z_L(t)} \sqrt{1 + \left( \frac{\omega(z,t)}{\partial z} \right)^2} \, dz = L \qquad (72)$$

The source of the above relation is that we know the arc length of IPMC is always constant and equal to *L*, and the above formulation is this equilibrium relationship, and the integration term is the arc length of IPMC that we have modeled it.

***Evidence 4:*** As you can follow in equations (70) to (72), it was shown that the common method to find IPMC tip displacement from $\omega(z,t)$ (Eq (70)) is absolutely wrong and it is just an approximation in the very small deformation situation. But using our proposed relation (Eq (71)) we can find accurate tip displacement of IPMC over time in all deformation situations (small to large deformation situations). This evidence also tells us again that our method is more complete and more accurate approach than previous models.

## III. PARAMETERS ESTIMATION AND MODEL VALIDATION

In this part, we want to validate our claim to find an accurate, well-defined relationship between input applied voltage ($V_I(t)$) and output tip displacement of IPMC ($\delta_{Tip}(t)$). In this direction, we have used experimental data that we have collected from IPMC. The method of measuring tip displacement is based on camera and using image processing algorithms where the detail of this system and algorithm and also details of required hardware apparatus have been described in our previous papers [2-5], and we don't talk about them here.

### A. Specification

About our sample, we should say that its dimension is almost $28 \times 6 \times 0.2$ mm$^3$; also the electrolyte is Na$^{+}$ and the type of membrane is DuPont™ Nafion 117 PFS that coated with two thin layers of Pt. In this model and basically in all physics-based models we have some specified physical parameters and some unknown parameters. The specified physical parameters have been reported in the references and have been collected in Table.1, but unknown parameters should be estimated in parameter estimation stage.

**Table. 1.** *Specified physical parameters*

| | |
|---|---|
| $F = 96458 \left[ \frac{C}{mol} \right]$ | $\kappa_e = 1.34 \times 10^{-6} \left[ \frac{F}{m} \right]$ |
| $Y = 5.71 \times 10^8 \, [Pa]$ | $C^- = 1200 \left[ \frac{mol}{m^3} \right]$ |
| $d = 1.03 \times 10^{-11} \left[ \frac{m^2}{s} \right]$ | $R = 8.3143 \left[ \frac{J}{mol.K} \right]$ |
| $T = 293 [K]$ | $L \times W \times h = 28 \times 6 \times 0.0889 \, [mm^3]$ |

### B. Parameter estimation and model validation

The method of validation is that we test the IPMC for three signals as input voltages, i.e., sine wave, chirp, and PRBS. That means we estimate the unknown parameters using half of each of these signals and test it with the second half. For example, if we measure the tip displacement of IPMC for 80 sec in response an arbitrary input voltage, the first we estimate the unknown parameters using the first 40 sec of the dataset and then we use the estimated parameters for the model and test it using the second 40 second of the dataset. The method of estimation is not the goal of this research, and we can find a variety of classic and intelligent methods to estimate unknown parameters that here we have used a genetic algorithm-based optimization method to find unknown parameters. The estimated parameters using genetic algorithms have been collected in Table.2.

**Table. 2.** Estimated parameters using a genetic algorithm

| $k_f$ | 0.5 | $\hat{Y}_6$ | $5 \times 10^{-3}$ |
|---|---|---|---|
| $\tau_f$ | 44 | $r_M$ | 0.15 |
| $k_g$ | 4.5 | $r_{ew}$ | 0.175 |
| $\tau_g$ | 0.07 | $r_{el}$ | 0.11 |
| $\hat{Y}_2$ | 0.04 | $\mu$ | 2.4 |
| $\hat{Y}_5$ | 0.03 | $\sigma_0$ | 0.10 |
| | $B_{mnp}$ | 0.003 | |

As you can see in Figs 4 to 6, this model is accurate enough, and it can follow the actual output precisely. But we need to assess the accuracy of model numerically that as a proper well known numerical measure we can propose the Normalized Mean Square Error (NMSE). The NMSE of our model in response to Sine wave, PRBS, and Chirp input signals are 0.07, 0.025, and 0.0047, respectively, and their average is 0.0332 that these numbers tell us proposed model is accurate enough and it can work properly even in large deformation situations and practical applications.

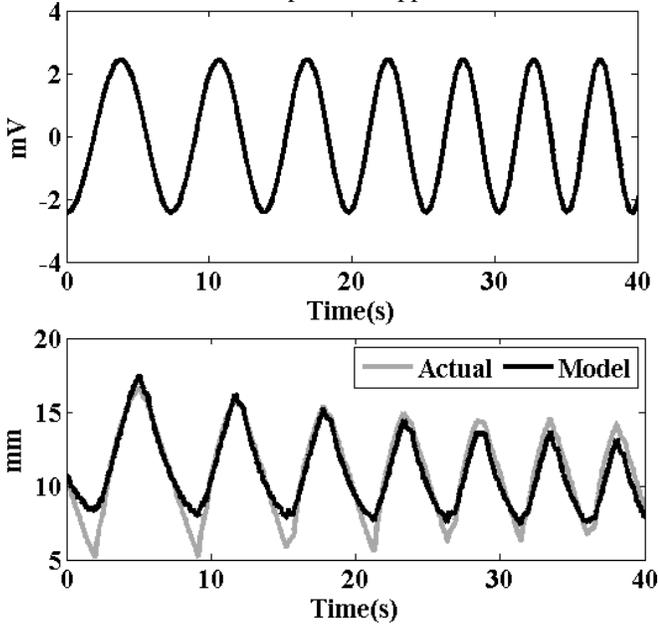

**Fig. 4.** Model validation. **(Up)** applied Chirp voltage to the IPMC (with a peak voltage of 2.3 V). **(Down)** the actual and estimated tip displacement of IPMC in $x$-direction axis in response to the Chirp input voltage.

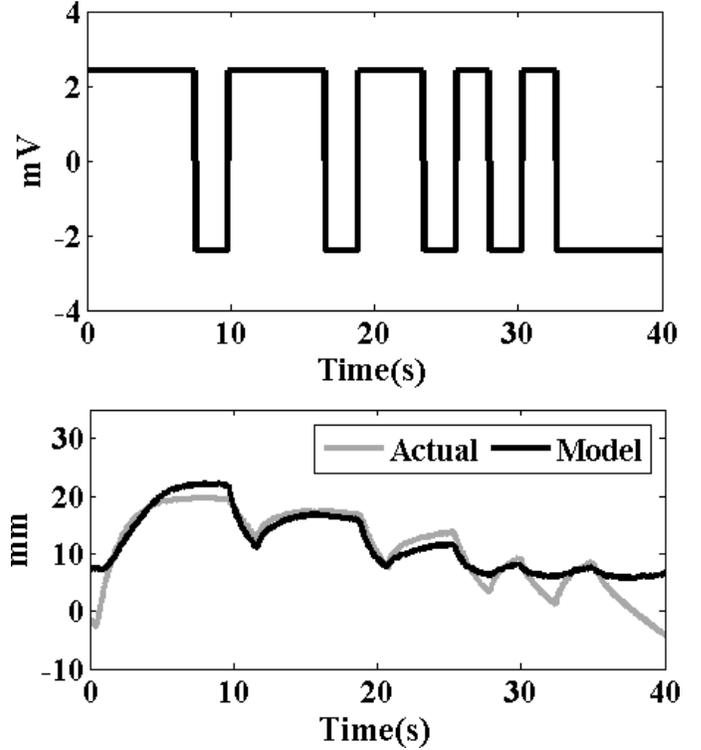

**Fig. 5.** Model validation. **(Up)** applied PRBS voltage to the IPMC (with a peak voltage of 2.3 V). **(Down)** the actual and estimated tip displacement of IPMC in the $x$-direction axis in response to the PRBS input voltage.

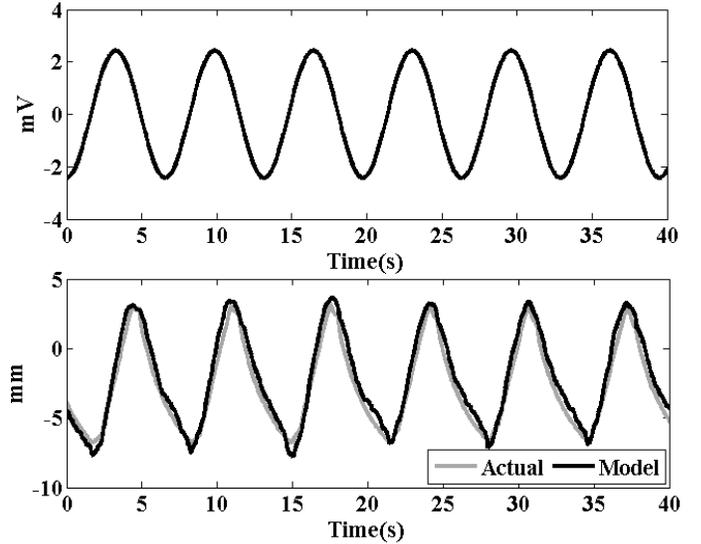

**Fig. 6.** Model validation. **(Up)** applied Sine wave voltage to the IPMC (with a peak voltage of 2.3 V and a frequency of 0.16 Hz). **(Down)** the actual and estimated tip displacement of IPMC in $x$-direction axis in response to the Sinewave input voltage.

## IV. CONCLUSION

It this paper as the first time we proposed a fully analytical and physics-based ion transport 3D and non-Linear model for large deformable IPMC. In this direction, based on three dimensional Nernst-Plank PDE, we found a well-defined and valid relationship between the input voltage and output tip displacement of IPMC that it is valid for large deformation situation. Also, using four provable pieces of evidence we explained that proposed model had chosen a proper way and it is more complete than the previous benchmark and well-known physics-based models. And finally, with some experimental examples, we showed that our model is accurate enough and is valid for large deformation situations.